\documentclass[12pt]{article}

\begin{document}

\title{The author replies}
\author{Michael Creutz\\
Physics Department, Brookhaven National Laboratory\\
Upton, NY 11973, USA
}
\maketitle

\begin{abstract}
I respond to the Bernard et al. comment on my letter ``Chiral
anomalies and rooted staggered fermions.''
\end{abstract}

In their comment on my letter ``Chiral anomalies and rooted staggered
fermions,'' \cite{Creutz:2007yg} Bernard, Golterman, Shamir, and
Sharpe \cite{bernard} do not address the main point that the chiral
symmetry group for the rooted theory has a higher rank than the target
theory.  This immediately calls into question whether the theories are
in the same universality class.  This misunderstanding is clear in
their item 4 when they say ``This kind of phenomenon \ldots is common
whenever the lattice theory has less symmetry than the continuum
theory.''  The issue is the reverse: the lattice theory has too much
symmetry.  To have a symmetry suddenly disappear in the continuum
limit is certainly not common.

Much of their discussion concerns whether ``taste'' symmetry is
restored.  Indeed, if the unrooted staggered theory did reduce to four
uncoupled but equivalent fermions in the continuum limit, one might
expect rooting to work.  This is especially true in perturbation
theory, where taking the fourth root of the determinant just
multiplies all fermion loops by one quarter.

But taste restoration is a considerably more complicated issue when
non-perturbative effects are taken into account.  Fermion doubling
generically arises from momenta in various corners of the Brillouin
zone.  These corners divide such that the various tastes appear with
differing physical chirality, i.e. their low energy modes use gamma
matrices that differ by signs.  The result is that the exact chiral
symmetry of staggered fermions represents a non-singlet symmetry
amongst the tastes.  The staggered determinant, even in the continuum
limit, does not correspond to the fourth power of a single fermion
theory.  The problem with rooting appears because the procedure
effectively averages over the different chiralities.  This is
inconsistent with the index theorem that says the one-flavor theory
should have a zero mode of a single chirality when the gauge field has
non vanishing winding.  The Bernard et al. comparison of rooted
staggered and overlap theories (in the paragraph following their
equation (5)) is misleading in this respect because four flavors of
overlap or Wilson fermions are forced by construction to have all
eigenvalues in identical quartets with identical chirality.

As mentioned in my letter, the eigenvalue matching that numerically
suggests taste symmetry restoration, such as seen in
Ref.~\cite{Follana:2005km}, must break during transitions between
topological sectors.  On passing from zero to unit winding, the four
nearly zero eigenmodes must have evolved from two eigenvalues dropping
down in the complex plane from above and two symmetrically rising from
below.

The underlying issue lies in the structure of the `t Hooft vertex
\cite{'tHooft:fv}, which the algorithm does not treat properly.
Before rooting this is a multilinear fermionic operator that strongly
couples all tastes.  To obtain the one flavor theory this should be
converted to a simple fermion bilinear.  However the exact chiral
symmetries of the unrooted theory are retained in the rooting process
and forbid the appearance of the correct target form.

The numerous numerical studies by the staggered community have shown
that rooted staggered quarks can quite accurately describe many
physical processes where the 't Hooft vertex does not play a major
role.  But any processes where these non-perturbative effects are
important will not be reproduced correctly by the algorithm.

\section*{Acknowledgments}
This manuscript has been authored under contract number
DE-AC02-98CH10886 with the U.S.~Department of Energy.  Accordingly,
the U.S. Government retains a non-exclusive, royalty-free license to
publish or reproduce the published form of this contribution, or allow
others to do so, for U.S.~Government purposes.

\end{document}